\documentclass[aps,amssymb,amsmath,superscriptaddress,showkeys,prl, twocolumn]{revtex4-1}
\usepackage[utf8x]{inputenc}
\usepackage{graphicx}
\usepackage{epsf}
\usepackage{natbib}
\usepackage{units}
\usepackage{ulem}

\usepackage{graphicx}% Include figure files
\usepackage{units}
\usepackage[mediumspace,mediumqspace,squaren]{SIunits}

\begin{document}

\title{Scanning Quantum Dot Microscopy}

\author{Christian Wagner}
\affiliation{Peter Gr{\"u}nberg Institut (PGI-3),Forschungszentrum J{\"u}lich, 52425 J{\"u}lich, Germany}
\affiliation{ J{\"u}lich Aachen Research Alliance (JARA)--Fundamentals of Future Information Technology, 52425 J{\"u}lich, Germany}
\author{Matthew F.~B.~Green}
\affiliation{Peter Gr{\"u}nberg Institut (PGI-3),Forschungszentrum J{\"u}lich, 52425 J{\"u}lich, Germany}
\affiliation{ J{\"u}lich Aachen Research Alliance (JARA)--Fundamentals of Future Information Technology, 52425 J{\"u}lich, Germany}
\author{Phillipp Leinen}
\affiliation{Peter Gr{\"u}nberg Institut (PGI-3),Forschungszentrum J{\"u}lich, 52425 J{\"u}lich, Germany}
\affiliation{ J{\"u}lich Aachen Research Alliance (JARA)--Fundamentals of Future Information Technology, 52425 J{\"u}lich, Germany}
\author{Thorsten Deilmann}
\affiliation{Institut f\"ur Festk\"orpertheorie, Westf\"alische Wilhelms-Universit\"at M\"unster, 48149 M\"unster, Germany}
\author{Peter Kr\"uger}
\affiliation{Institut f\"ur Festk\"orpertheorie, Westf\"alische Wilhelms-Universit\"at M\"unster, 48149 M\"unster, Germany}
\author{Michael Rohlfing}
\affiliation{Institut f\"ur Festk\"orpertheorie, Westf\"alische Wilhelms-Universit\"at M\"unster, 48149 M\"unster, Germany}
\author{Ruslan Temirov}
\email[corresponding author: ]{r.temirov@fz-juelich.de}
\affiliation{Peter Gr{\"u}nberg Institut (PGI-3),Forschungszentrum J{\"u}lich, 52425 J{\"u}lich, Germany}
\affiliation{ J{\"u}lich Aachen Research Alliance (JARA)--Fundamentals of Future Information Technology, 52425 J{\"u}lich, Germany}
\author{F. Stefan Tautz}
\affiliation{Peter Gr{\"u}nberg Institut (PGI-3),Forschungszentrum J{\"u}lich, 52425 J{\"u}lich, Germany}
\affiliation{ J{\"u}lich Aachen Research Alliance (JARA)--Fundamentals of Future Information Technology, 52425 J{\"u}lich, Germany}

%\abbreviations{STM,AFM,PP}
\keywords{}

\begin{abstract}

Interactions between atomic and molecular objects are to a large extent defined by the nanoscale electrostatic potentials which these objects produce. We introduce a scanning probe technique that enables three-dimensional imaging of local electrostatic potential fields with sub-nanometer resolution. Registering single electron charging events of a molecular quantum dot attached to the tip of a (qPlus tuning fork) atomic force microscope operated at 5 K, we quantitatively measure the quadrupole field of a single molecule and the dipole field of a single metal adatom, both adsorbed on a clean metal surface. Because of its high sensitivity, the technique can record electrostatic potentials at large distances from their sources, which above all will help to image complex samples with increased surface roughness.

\end{abstract}

\maketitle

The atomic structure of matter inevitably leads to local electrostatic fields in the vicinity of nanoscale objects even if they are neutral \cite{Smoluchowski1941}. For this reason electrostatic forces often dominate the interactions between nanostructures. In spite of their omnipresence, experimental access to such local electrostatic fields is a formidable challenge, Kelvin probe force microscopy (KPFM) being the most promising attempt to address it so far \cite{Nonnenmacher1991,Gross2009,Gross2012}. However, since KPFM measures the contact potential difference between surfaces, which by definition are extended objects, it inevitably involves considerable lateral averaging, especially for larger probe-to-surface distances. True three-dimensional imaging of local electrostatic fields in a broad distance range is therefore difficult with KPFM \cite{Schuler2014}. 

Here we introduce a scanning probe technique that provides a contact-free measurement of the electrostatic potential in all three spatial dimensions, without the drawback of distance dependent averaging. This is possible because the method, unlike KPFM, directly probes the local electrostatic potential at a well-defined sub-nanometer-sized spot in the junction. Besides its high spatial resolution, our technique benefits from a remarkable sensitivity that allows, e.g., the detection and quantitative evaluation of the electrostatic potential $7~\rm{nm}$ above a single adatom on a metal surface.

\begin{figure}
\centering
\includegraphics[width=8cm]{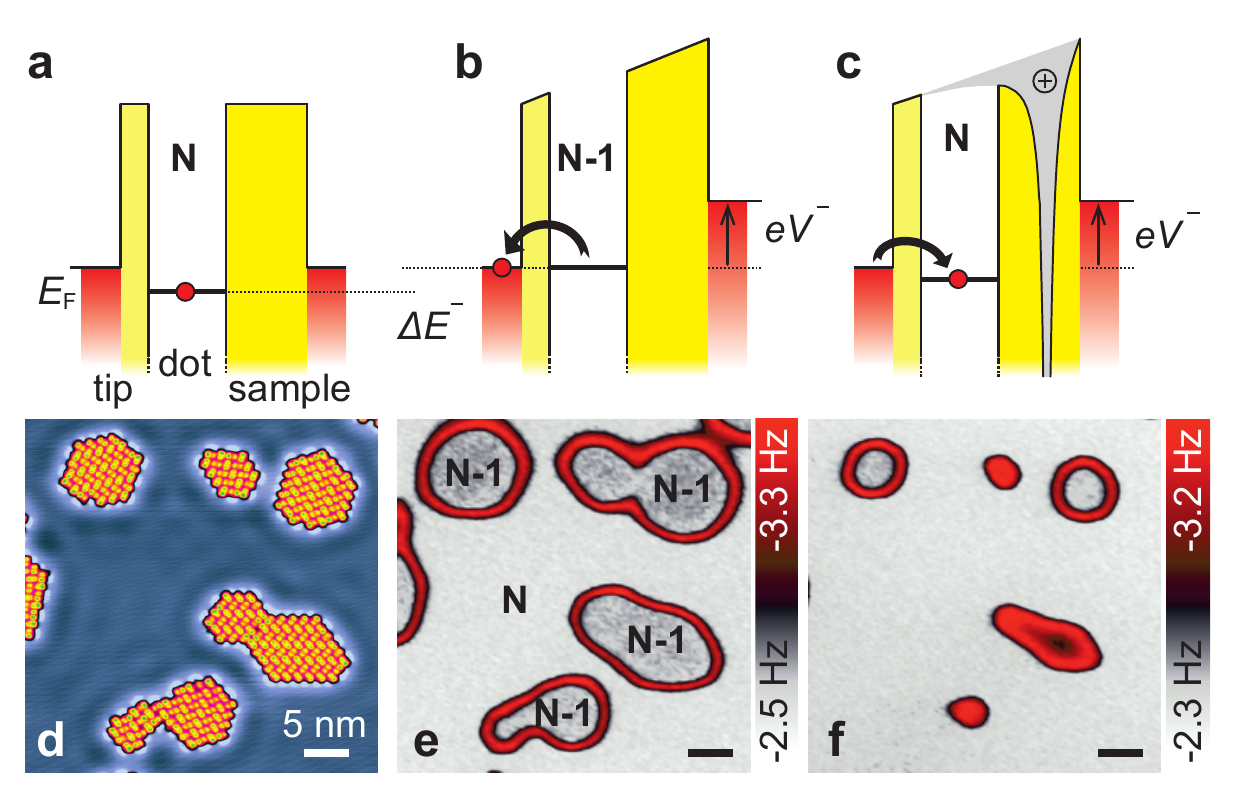}
\caption{Working principle of SQDM. (\textbf{a-c}) Energy diagrams showing the QD attached to a scanning probe tip. (\textbf{a}) In the absence of a sample bias an exemplary level of the QD is occupied (QD charge state N). (\textbf{b}) When a critical sample bias $V^-$ is reached, one electron tunnels from the QD into the tip (QD charge state N-1). (\textbf{c}) If a local charge at the surface modifies the potential in the junction, the QD level shifts and becomes re-occupied (QD charge state N). (\textbf{d}) STM image of PTCDA islands on Ag(111). Here and on all further images a $5~\rm{nm}$ scale bar is shown. (\textbf{e}) Constant height raw $\Delta f$ image of the area in (d) recorded at $z_\mathrm{tip}=3~\rm{nm}$ and $V= -990~\rm{mV}$ (marked in Fig.~\ref{fig2}b). Prominently visible in red are the locations where $V= V^-$. These are electrostatic equipotential lines. The charge state of the QD in the different regions is labelled according to (a) and (c). (\textbf{f}) Same as (e), but recorded at $V= -910~\rm{mV}$.}
\label{fig1} 
\end{figure} 

We image the electrostatic potential using a nanometer-sized quantum dot (QD) attached to the apex of the scanning probe tip (Fig.~1a). The tunneling barrier between the QD and the tip is sufficiently large so that the electronic levels of the QD experience only weak hybridization \cite{Level_width}. In the experiment the electronic levels of the QD are gated with respect to the Fermi level $E_{\rm{F}}$ of the tip by applying a bias voltage to the tip-surface junction (Fig.~1b) \cite{Woodside2002, Stomp2005, Cockins2010, Fernandez-Torrente2012,Lotze2013}. In this way, the charge state of the QD can be changed, e.g.~if the bias voltage $V$ applied to the junction reaches a critical value $V^-$ that aligns one of the QD's occupied electronic levels with $E_{\rm{F}}$, thus inducing its depopulation (Fig.~1b). With this device the measurement of a local electrostatic potential field $\Phi(x,y,z)$, caused e.g. by a surface adsorbate, is possible because the electronic levels of the QD shift in response to any perturbation of the potential at the position $(x,y,z)$ of the QD. Although small, these shifts can be detected by their effect on the charge state, if an occupied or empty level, gated by the bias voltage, lies in the close vicinity of $E_{\rm{F}}$ (Fig.~1c). In essence, detecting charging events of the QD while scanning the three-dimensional half-space above the surface is the core working principle of our method, to which we refer to as \textit{scanning quantum dot microscopy} (SQDM).

Figs.~1e-f illustrate the outcome of SQDM imaging, by visualizing the effect which monolayer-thick islands of perylene tetracarboxylic dianhydride (PTCDA) adsorbed at the Ag(111) surface have on the charge state of the QD when the latter is scanned at a distance of $\approx 3~\rm{nm}$ across the surface. The red contours in Figs.~1e-f mark locations where the QD changes its charge state. Note that these contours follow the shape of the standing wave pattern (Fig.~1d) which is formed by the surface state as it is scattered by the perturbed electrostatic potential in the surface \cite{Crommie1993}. This is an initial indication that the gated QD is indeed sensitive to the electrostatic potential created by the sample in the half-space above it. In the remainder of the paper we present experimental results that unambiguously confirm this conjecture. 

\begin{figure}
\centering
\includegraphics[width=8 cm]{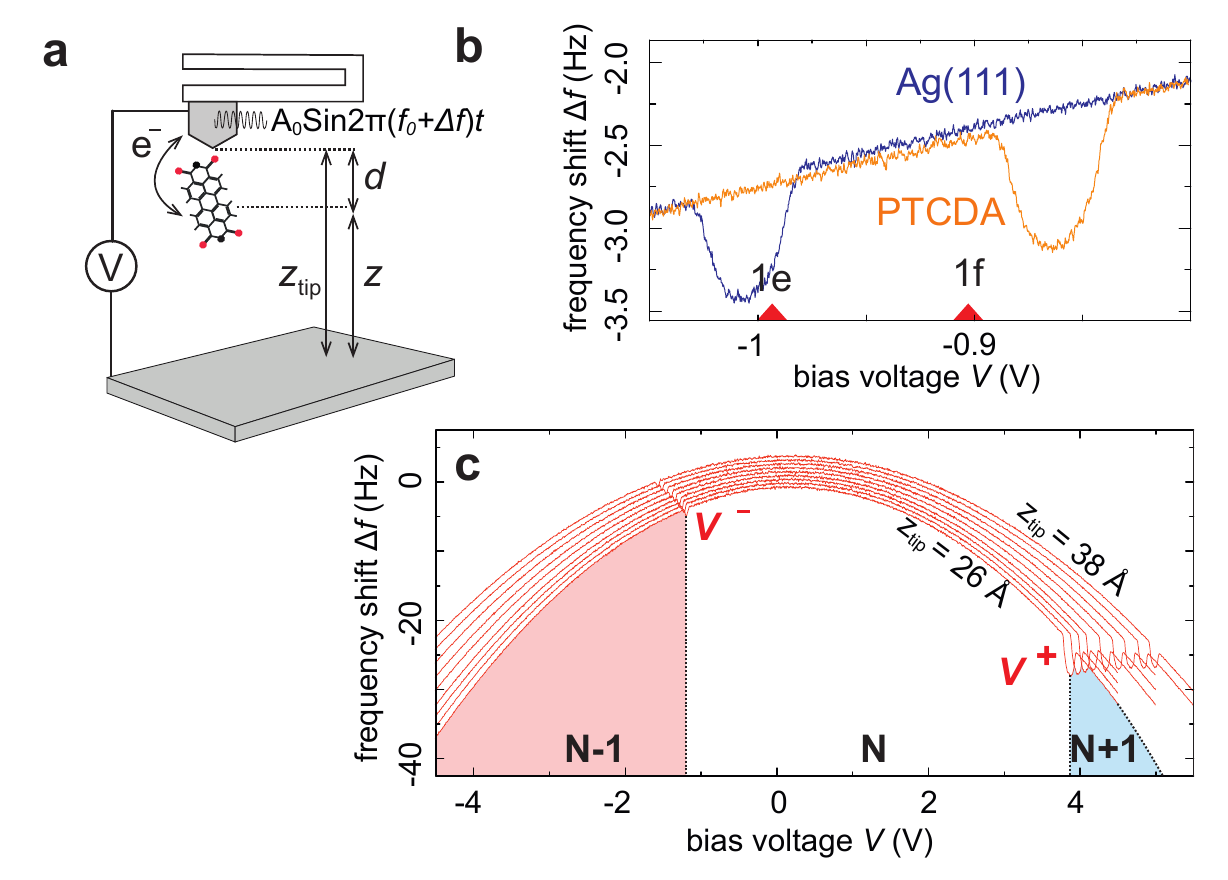}
\caption{SQDM implementation combining a molecular QD and a NC-AFM. (\textbf{a}) Schematic view of the QD sensor: A single PTCDA molecule is chemically bonded to the AFM tip via a corner oxygen atom. The definitions of $z_\mathrm{tip}$ and $d$ are indicated. The calibration of $z_\mathrm{tip}$ was performed as described in references \cite{Wagner2012,Wagner2014}  (\textbf{b}) $\Delta f(V)$ spectra taken above the clean Ag(111) surface (blue) or above a PTCDA island (orange). The center of each dip determines $V^-$. The voltages $V$ used for scanning Figs.~\ref{fig1}e and f are indicated with red triangles. (\textbf{c}) Series of $\Delta f(V)$ spectra, recorded at varying $z_{\rm{tip}}$ above the bare Ag(111) surface (consecutive curves are offset by $0.5~\rm{Hz}$). For the curve recorded at $z_\mathrm{tip}=~26~\rm{\AA}$, the charge states of the QD are indicated by the color shading and the total number of electrons, $N-1$, $N$ and $N+1$.}
\label{fig2} 
\end{figure} 

First we describe the structure of the junction and details of the measurement protocol. In our realization of SQDM, the role of the QD is played by a single PTCDA molecule which is connected to the tip of a commercial qPlus tuning fork \cite{Giessibl2003} non-contact atomic force / scanning tunneling microscope (NC-AFM/STM) from CREATEC, operated at $5~\rm{K}$ and in ultra-high vacuum (Fig.~2a). The molecule is attached to the tip through a single chemical bond between the outermost atom of the tip and one of the corner oxygen atoms of the PTCDA, using a well-described manipulation routine \cite{Wagner2012,Wagner2014} that in brief is described as follows: Firstly the isolated PTCDA molecule adsorbed on Ag(111) surface is approached by the silver-covered AFM/STM tip directly above one of its corner oxygen atoms. These atoms are known to show reactivity towards silver. At a tip-surface distance of about $z_{\rm{tip}}$=6.5\AA~the chosen oxygen atom jumps up to establish a chemical bond to the tip. By this tip-oxygen bond the entire PTCDA molecule can be lifted off the surface. As the final bond between the molecule and the surface is broken, the attractive interaction with the surface aligns the molecule along the axis of the tip \cite{Wagner2012,Wagner2014}, leaving it in a configuration that is suitable for SQDM (Fig.~2a). Scanning the tip at sufficiently large distances from the surface ensures that the PTCDA QD does not change its configuration on the tip during the SQDM experiment. For a given distance between tip and sample $(z_{\rm{tip}})$ the quantum dot is therefore always located at coordinate $z$, where $d=z_{\rm{tip}}-z$ is the distance of the quantum dot from the tip apex (Fig.~2a).

Since electrostatic potential measurements in SQDM are based on changes of the QD's electron occupation, a sensitive detection of charging events is crucial. In the present realization of SQDM this is accomplished by registering abrupt steps in the tip-sample force which always accompany the change of the QD's charge state \cite{Woodside2002, Stomp2005, Cockins2010, Lotze2013}. In the qPlus NC-AFM, these steps show up as sharp dips in the frequency shift curve $\Delta f(V)$ \cite{Giessibl2003, Measurement} (Fig.~2b). The positions of the sharp $\Delta f$ features on the bias voltage axis are the principal signal which is evaluated in SQDM.

\begin{figure}
\centering
\includegraphics[width=8cm]{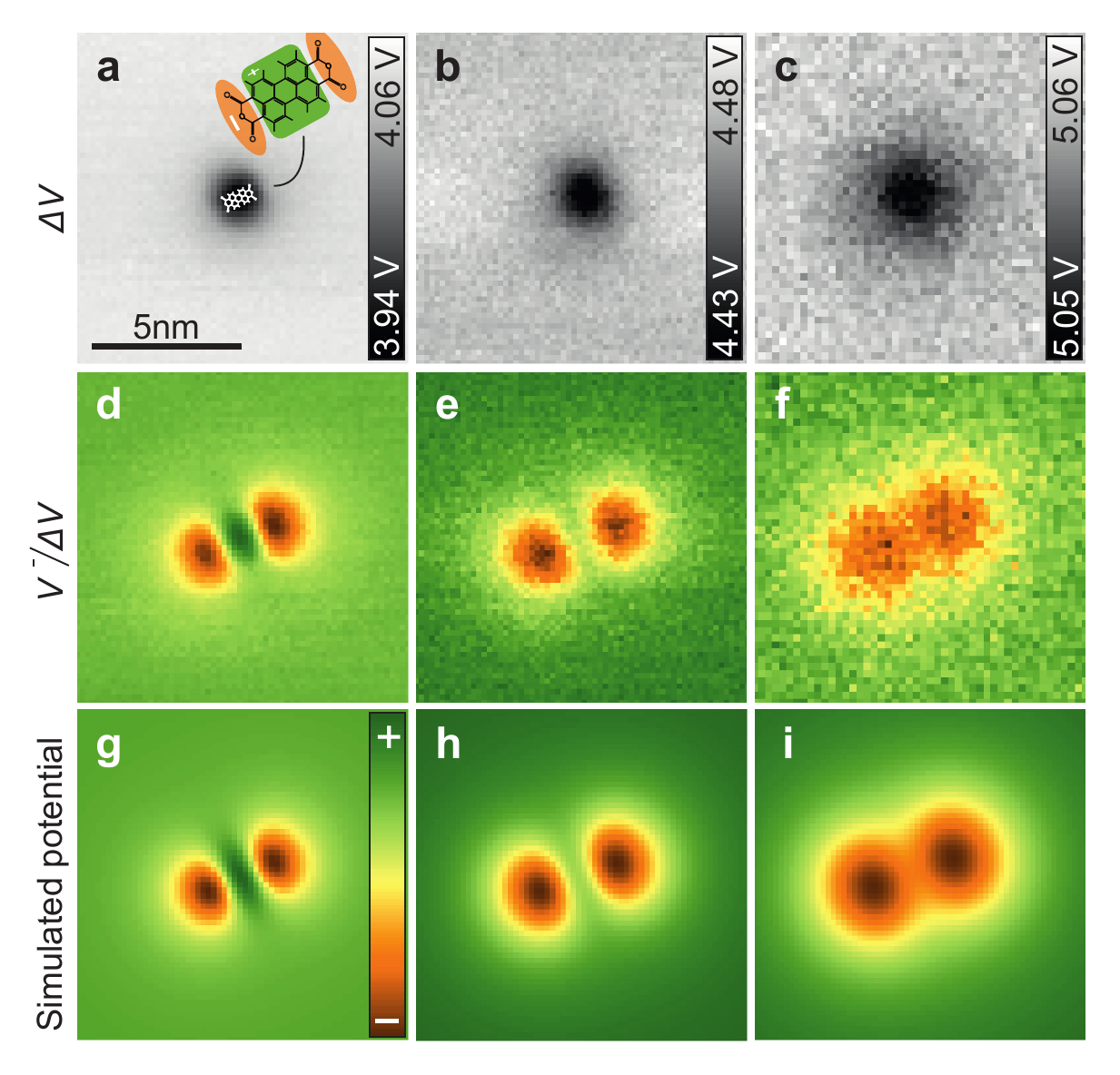}
\caption{PTCDA on Ag(111): Separation of gating efficiency contrast (which includes topographic contrast) and electrostatic potential contrast in SQDM. (\textbf{a-c}) Experimental $\Delta V(x,y)$ maps (related to gating efficiency, for a definition and more details cf.~text) recorded over an isolated PTCDA molecule adsorbed on the Ag(111) surface. Panel (a) contains  a scale drawing of the PTCDA molecule (white) and in the inset an enlarged structure formula, on which the quadrupolar charge distribution is indicated schematically. (\textbf{d-f}) Experimental $V^-/\Delta V(x,y)$ maps (related to electrostatic potential, for a definition and more details cf.~text) of the same area as in (a-c). Maps (a) and (d) were recorded at $z_{\rm{tip}}=24~\rm{\AA}$, (b) and (e) at $28~\rm{\AA}$, (c) and (f) at $36~\rm{\AA}$ (\textbf{g-i}) Simulated electrostatic potential of adsorbed PTCDA at $z=16~\rm{\AA}$ (g), $22~\rm{\AA}$ (h), and $28~\rm{\AA}$ (i). The color scales in (d-i) were adjusted to facilitate the contrast of each figure.}
\label{fig3} 
\end{figure}

A slight complication in the measurement of local electrostatic potential fields by SQDM arises from the fact that topographic features in the sample surface can also lead to changes in the QD's charge state. Since nanostructures which produce local electrostatic potential fields usually have topographic signatures, both influences have to be disentangled from one another. As it turns out, the simultaneous analysis of \textit{two} levels of the QD offers a straightforward possibility to achieve this. Therefore, in the experiments to be discussed below, we access two electronic levels of the molecular QD: one empty, the other occupied by one electron. To this end, the data in Fig.~2c, measured above the bare Ag(111) surface, exhibits two charging events: gating the occupied (empty) level to $E_{\rm{F}}$ reduces (increases) the charge by one electron at $V^-$ ($V^+$). We note in passing that Fig.~2c also reveals that both charging events appear on top of the well-known parabola which originates from the attractive interaction between the opposing electrodes of the biased tip-surface junction \cite{Gross2009}. 

The fact that topographic signatures can charge or discharge the QD if the tip is scanned across the surface in constant-height mode, i.e.~at a fixed $z$ (see Fig.~\ref{fig2}a), is naturally explained by changes of the junction capacitance with the distance between tip and sample \cite{Stomp2005}. The effect is illustrated in Fig.~\ref{fig2}c by the observation that the absolute values $|V^+|$ and $|V^-|$ increase with the distance $z$ between the tip and the bare Ag(111) surface. Here we describe this behaviour in terms of a quantity $\alpha$ that we call gating efficiency. A smaller value of $\alpha$ implies that a larger bias is needed to align any given PTCDA level with $E_{\rm{F}}$. A larger $z$ thus goes along with a smaller gating efficiency. In fact, at a fixed $z$, the quantity $\Delta V \equiv V^+-V^-$ is inversely proportional to the gating efficiency $\alpha$: $\alpha = C/\Delta V$ where $C$ is a constant \cite{support}. 

In contrast to $\alpha$, a local electrostatic potential $\Phi$ at the position of the QD shifts $V^-$ and $V^+$ rigidly on the voltage axis. For a fixed $z$, the separation of $\Phi^*$ from topography, dielectric contrast and all other factors that influence the gating efficiency can be achieved straightforwardly, by $\Phi^*= -\alpha V^-+\Phi^*_0=-CV^-/\Delta V+\Phi^*_0$ \cite{support}, where $\Phi^*_0$ is a constant. Note that $\Phi^*$ is the local electrostatic potential created by the nanostructure in the presence of the metallic tip. Its relation to the electrostatic potential $\Phi$ of the nanostructure in the absence of the tip will be discussed below.  

We are now in the position to demonstrate the power of SQDM by mapping out the local electrostatic potential field of a nanostructure. As the latter we choose an individual PTCDA molecule adsorbed on the Ag(111) surface. Its field is expected to contain two major contributions, a quadrupolar field that is produced by the internal charge distribution of the molecule (negative partial charges at the oxygen atoms, see \ref{fig3}a) and a dipolar field due to the well-known electron transfer from Ag(111) to PTCDA upon adsorption \cite{Zou2006}. The experimental quantities  $\Delta V(x,y)$, inversely proportional to the gating efficiency, and $V^-/\Delta V(x,y)$, proportional to the electrostatic potential (up to a constant offset), are displayed in Figs.~3a-f for $z_{\rm{tip}}=24~\rm{\AA},~28~\rm{\AA},~36~\rm{\AA}$.

A visual inspection of the images in Figs.~3d-f immediately indicates their close resemblance to the molecular quadrupole field. This is reinforced  by a comparison to the results of a microelectrostatic simulation, in which the internal charge distribution of PTCDA, as calculated by density functional theory \cite{Becke1993}, its screening by the metal, and a homogeneous charge transfer from the metal to the molecule have been taken into account. Since the precise amount of transferred charge is not known, this was treated as a fit parameter ($q=-0.09~\mathrm{e}$). The result is shown in Figs.~3g-i, which show excellent qualitative agreement with the corresponding images in Figs.~3d-f. Remarkably, if we compare the distances from the surface at which the model potential had to be calculated $(z=16~\rm{\AA},~22~\rm{\AA},~28~\rm{\AA})$  in order to reproduce the experimental images at $z_{\rm{tip}}=24~\rm{\AA},~28~\rm{\AA},~36~\rm{\AA}$, we obtain a systematic difference of $d=(7~\pm~1)~\rm{\AA}$ (see Fig.~\ref{fig2}a). This shows that the electrostatic potential is probed at a point approximately $7~\rm{\AA}$ below the tip apex, hence precisely at the position of the QD, as to be expected from the proposed mechanism of SQDM.

\begin{figure}
\centering
\includegraphics[width=8cm]{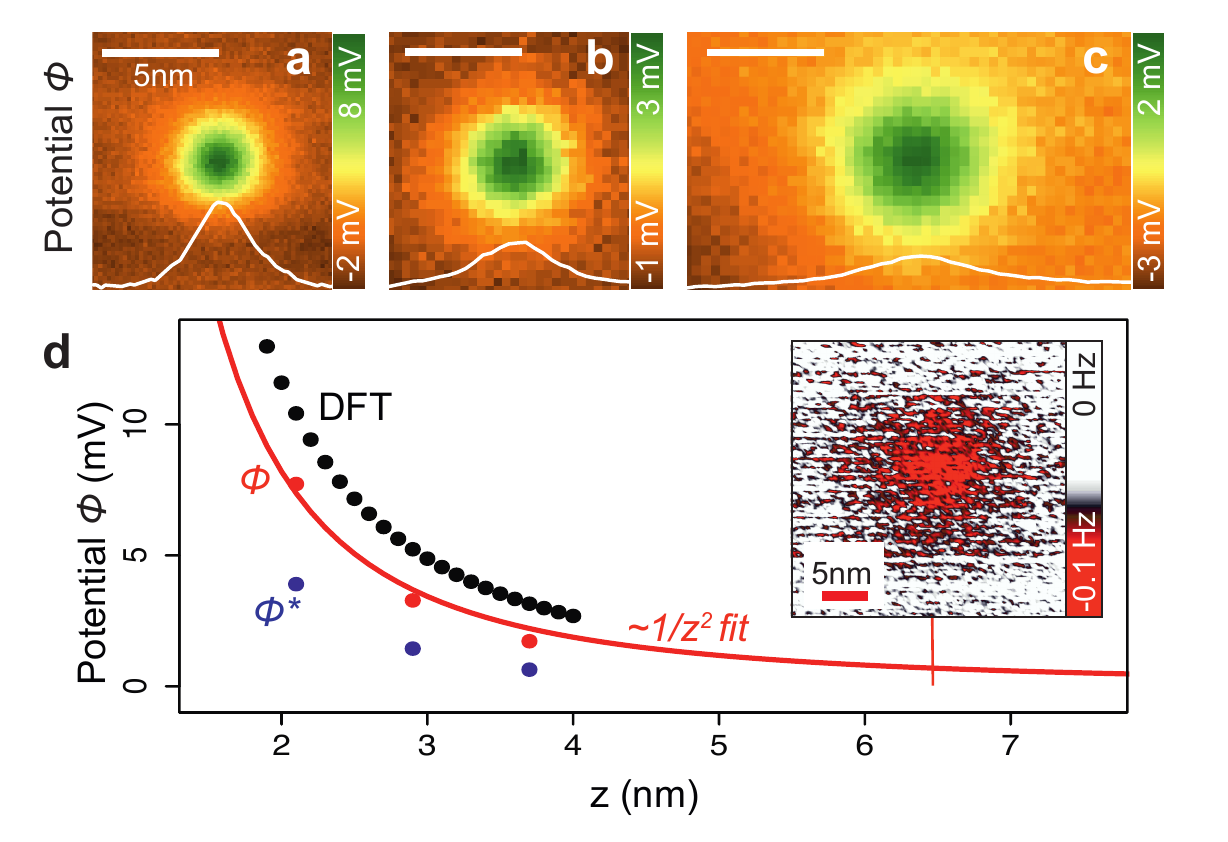}
\caption{Adatom on a surface: Quantitative electrostatic potential measurements with SQDM. (\textbf{a-c}) Electrostatic potential $\Phi$ maps measured above a silver adatom on Ag(111) at $z = 21~\rm{\AA}$ (A), $29~\rm{\AA}$ (b), and $37~\rm{\AA}$ (c). The scale bars show the absolute values of the electrostatic potential $\Phi$, obtained as described in the text. Line profiles through the adatom are shown in white. (\textbf{d}) Comparison of the experimental $\Phi^\star$ (blue), $\Phi$ (red) and DFT-calculated (black) potentials vertically above the adatom. Red line shows $1/z^2$ fit of the DFT data. For the experimental data $z$ is the distance between the point inside the QD at which the electrostatic potential is measured, and the surface. For DFT $z$ is the distance from the surface at which the potential of the adatom was calculated. \textbf{Inset}: Constant height raw $\Delta f$ image recorded at $z=6.3~\rm{nm}$ with an applied bias of $V=9.6~\rm{V}$, close to $V^+$ for this $z$.}
\label{fig4} 
\end{figure}

We then choose the well-known Smoluchowski dipole \cite{Smoluchowski1941}, created here by a single metal adatom on a metal surface (Figs.~4a-c), to demonstrate fully quantitative three-dimensional electrostatic potential imaging. For the latter it must be taken into account that the constants $C$ and $\Phi^*_0$ are $z$-dependent. This can be taken care of by performing $V^-_0(z)$ and $V^+_0(z)$ reference measurements for a fixed set of heights $z$ at a location where the local electrostatic potential $\Phi$ is taken to be zero, e.g.~above the bare Ag(111) surface. In this way $\Phi$ can be evaluated in all locations $(x,y,z)$ from $\Phi^*(x,y,z)=-\alpha_0(z) \left(\frac{V^-(x,y,z)}{\Delta V(x,y,z)}\Delta V_0(z)-V^-_0(z)\right)$ \cite{support}, where $\alpha_0(z)$ is the $z$-dependent gating efficiency when the QD-tip is above the bare Ag(111) surface. In the simplest case, $\alpha_0(z)=d/(z+d)$, if a plate capacitor geometry is assumed. Fig.~4d shows the experimental electrostatic potential vertically above the adatom, evaluated by the above formula, in comparison to the result of a DFT calculation \cite{DTFnote}. 

Before making the comparison one should note that DFT yields the electrostatic potential $\Phi$ in absence of the tip. It is also clear that the grounded tip screens the local electrostatic potential $\Phi$ to a smaller value $\Phi^*$. Taking into the account this screening \cite{screening}, we obtain the experimental $\Phi$ that comes at about 70\% of DFT values. We consider the observed agreement in magnitude a remarkable verification of the SQDM performance in quantitative mapping of the electrostatic potential. The remaining small discrepancy between the theory and the experiment can be explained by an effective increase of $\alpha$ (in comparison to the plate capacitor case) caused by the strong curvature of the sharp metal tip used in the experiment. We note that this influence can be quantified by measuring a structure whose electrostatic potential is known and then transferred to any other experiment with the same tip.  

Finally, we comment on the sensitivity of our electrostatic potential field measurement, again using the adatom as an example. Using the fact that the experimentally determined $\Phi(z)$  closely follows the $1/z^2$ behavior that is expected for a point dipole (see Fig. \ref{fig4}d) we find that the $\Phi(z)$ reaches the sensitivity limit defined by our bias voltage measurement resolution of $\sim 1~\rm{mV}$ at a distance of about $z=7~\rm{nm}$ from the surface. To confirm this estimate, the inset in Fig.~4d shows that the Smoluchowski dipole field of the adatom is indeed still detectable at a distance of $z=6.3~\rm{nm}$ from the surface.  
  
In conclusion, we have reported a scanning probe technique that is able to provide truly three-dimensional, so far elusive, maps of the electrostatic potential field with nanometer resolution. The current realization of scanning quantum dot microscopy (SQDM) is based on a single molecular quantum dot attached to the tip of a scanning probe microscope. We demonstrate the power of SQDM by measuring the electrostatic potential fields of the Smoluchowski dipole created by a single silver adatom and by  the quadrupole moment of a single PTCDA molecule. Since the quantum dot serves as a sensor of the electrostatic potential which at the same time transduces this signal to a charging event, the technique is a particularly fascinating variant of the general sensor/transducer concept for scanning probe microscopy introduced earlier \cite{Weiss2010, Chiang2014, Hapala2014,filteramplifier}. Here, however, the transduction involves electronic rather than the mechanical degrees of freedom that were utilized in previous work. As a consequence of its high sensitivity, SQDM can be applied to rough and high aspect ratio samples, opening the possibility to study, e.g., semiconductor devices and biological samples. Moreover, the combination of high sensitivity and spatial resolution suggests the possibility of reading nanoscale electric memory cells entirely contact- and current-free. We note in passing that beside AFM other detection schemes of the charging events are conceivable. Finally, we stress that a molecular quantum dot, although particularly attractive, does not exhaust all possibilities. SQDM probes with nano-fabricated quantum dots on standard silicon AFM cantilevers can be envisioned, extending the applicability beyond ultra-high vacuum and cryogenic temperatures.

The authors gratefully acknowledge the computing time granted by the John von Neumann Institute for Computing (NIC) and provided on the supercomputer JUROPA at J\"ulich Supercomputing Centre (JSC).

\widetext
\clearpage

\begin{center}
\textbf{\Large Scanning Quantum Dot Microscopy:\\
Supplemental Information}
\end{center}

%%%%%%%%%% Merge with supplemental materials %%%%%%%%%%
%%%%%%%%%% Prefix a "S" to all equations, figures, tables and reset the counter %%%%%%%%%%
\setcounter{equation}{0}
\setcounter{figure}{0}
\setcounter{table}{0}
\setcounter{page}{1}
\makeatletter
\renewcommand{\theequation}{S\arabic{equation}}
\renewcommand{\thefigure}{S\arabic{figure}}
\renewcommand{\bibnumfmt}[1]{[S#1]}
\renewcommand{\citenumfont}[1]{S#1}
%%%%%%%%%% Prefix a "S" to all equations, figures, tables and reset the counter %%%%%%%%%%

\clearpage

\begin{figure}\centering
\includegraphics[width=15cm]{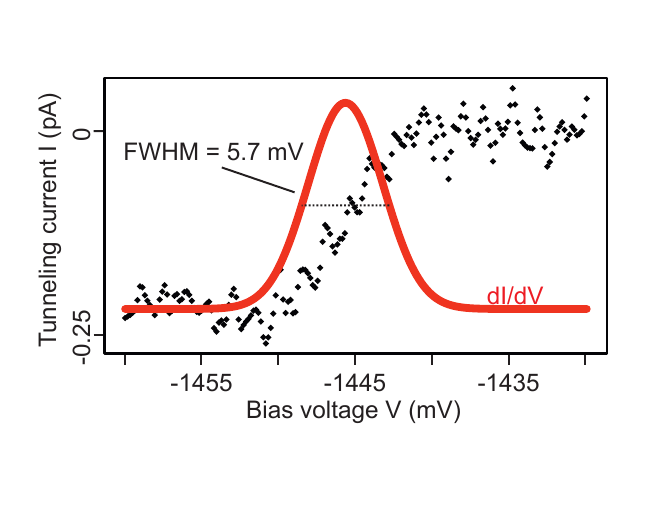}
\caption{\textbf{Interaction between the PTCDA quantum dot and the tip electrode}, as determined from charge transport between the tip and the Ag(111) surface through the PTCDA quantum dot. The black data points show a sharp increase of the tunneling current $I$ that occurs when the occupied level of the quantum dot crosses $E_F$ of the tip at the bias voltage $V^+$ and thus moves inside the bias window, becoming a channel for charge transport. The red curve shows the corresponding peak in the differential conductance $dI/dV$, the width of which is directly related to the width of the PTCDA level. Taking into account a thermal broadening of $5.4kT$, the Figure reveals that the broadening of the occupied PTCDA level through the interaction with the tip is smaller than 3.5 mV.
 }
\label{figS}
\end{figure} 

\clearpage
\section{Extraction of constant height $\Phi^*(x,y)$ and $\alpha(x,y)$ images}

\begin{figure}\centering
\includegraphics[width=15cm]{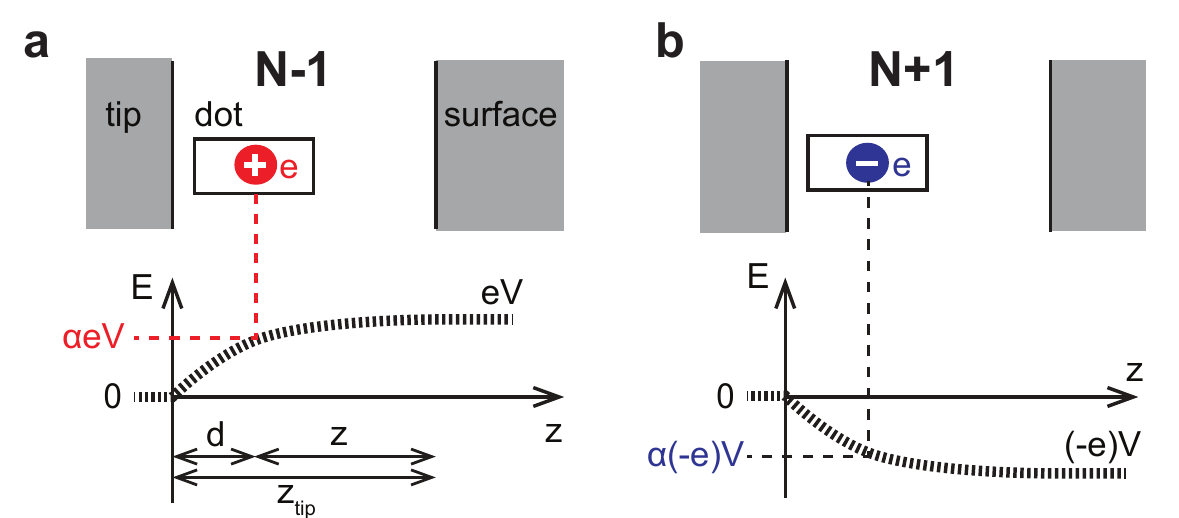}
\caption{\textbf{Schematic diagram of the SQDM junction.} In \textbf{a}, the dot is positively charged with one hole (total number of electrons $N-1$), while in \textbf{b} the dot is negatively charged with one electron (total number of electrons $N+1$). $V$ is the bias applied to the tip-surface junction. In both figures the bias $V$ is positive. The thick dotted lines shows the energy of a hole (panel \textbf{a}, charge $+e$) or the energy of an electron  (panel \textbf{b}, charge $-e$) as a function of $z$. The positive bias $V$ tends to stabilize the electron in the dot (panel \textbf{a}), but to destabilize the hole in the dot (panel \textbf{a}). The stabilization and destabilization energies are shown in blue and red, respectively.  }
\label{figSI0}
\end{figure} 

Considering the diagrams shown in Figs.~\ref{figSI0}a-b, the energies of the states of the junction can be written as follows:

\begin{equation} \label{eq:N-1}
E(N-1)=E_{\rm{hole}}+\alpha e V +e\Phi^*
\end{equation}

\noindent and 

\begin{equation} \label{eq:N+1}
E(N+1)=E_{\rm{el}}-\alpha e V-e\Phi^*,
\end{equation}

\noindent where $E_{\rm{hole}}$ ($E_{\rm{el}}$) is the energy needed to create a hole (electron) in the quantum dot when no bias voltage $V$ is applied to the junction. $\pm\alpha e V$ is the energy associated with the position of the hole or electron in the electrostatic potential created by the bias voltage $V$. $\alpha $ is the gating efficiency which determines which fraction of the bias voltage $V$ drops between the tip and the quantum dot. Finally, $\Phi^*$ is an additional electrostatic potential present at the position of the quantum dot, created, e.g., by a nanostructure in the vicinity. In our experiment $\Phi^*$ is the measured quantity.

The charging conditions can be written as $E(N-1)=0$ and $E(N+1)=0$. This leads to a pair of equations

\begin{equation} \label{eq:N-1=0}
E_{\rm{hole}}+\alpha e V^- +e\Phi^*=0
\end{equation}

\begin{equation} \label{eq:N+1=0}
E_{\rm{el}}-\alpha e V^+-e\Phi^*=0,
\end{equation}

\noindent in which $ V^+$ and $V^-$ are the charging voltages measured in the experiment (cf.~main text). From eqs.~\ref{eq:N-1=0} and \ref{eq:N+1=0} we obtain 

\begin{equation} \label{eq:alpha}
\alpha=\frac{E_{\rm{hole}}+E_{\rm{el}}}{e(V^+-V^-)}=\frac{E_{\rm{hole}}+E_{\rm{el}}}{e\Delta V}
\end{equation}

\begin{equation} \label{eq:phi}
\Phi^*=-\frac{V^-}{e\Delta V}(E_{\rm{hole}}+E_{\rm{el}}) - \frac{E_{\rm{hole}}}{e}
\end{equation}

\noindent Assuming that neither $E_{\rm{hole}}$ nor $E_{\rm{el}}$ changes when the tip is scanned at constant height (i.e.~fixed $z$) across the surface, eqs.~\ref{eq:alpha} and \ref{eq:phi} show that from the measured $V^+(x,y)$, $V^-(x,y)$ at a given $z$ we can obtain maps of the gating efficiency $\alpha (x,y)$ and the potential $\Phi^*(x,y)$ at this $z$, up to a scaling factor and an offset. In Figs.~3a-c of the main paper we plot the measured $\Delta V$, related to $\alpha ^{-1}$, and in Figs.~3d-f of the main paper we plot the measured dimensionless quantity $V^-/\Delta V$, related to $\Phi^*$.

\section{Reference Measurement at constant $z$}

The unknown scaling factor $(E_{\rm{hole}}+E_{\rm{el}})/e$ (appearing in the main text as $C$) and offset $E_{\rm{hole}}/e$ (appearing in the main text as $\Phi^*_0$) in eqs.~\ref{eq:alpha} and \ref{eq:phi} can be eliminated by a reference measurement at a point $(x_0,y_0)$ at which the local electrostatic potential $\Phi^*$ is zero, e.g.~above the bare Ag(111) surface. It is important that this reference measurement is carried out at the same $z$ at which $\alpha (x,y)$ and $\Phi^*(x,y)$ are further evaluated. If the local electrostatic potential $\Phi^*$ is zero, eqs.~\ref{eq:N-1=0} and \ref{eq:N+1=0} become

\begin{equation} \label{eq:cal1}
E_{\rm{hole}}=-\alpha_0 eV^-_{\rm{0}}
\end{equation}

\begin{equation} \label{eq:cal2}
E_{\rm{el}}=\alpha_0 eV^+_{\rm{0}}
\end{equation}

\noindent where $\alpha_0=\alpha(x_0,y_0)$, $V^-_{\rm{0}}=V^-(x_0,y_0)$ and $V^+_{\rm{0}}=V^+(x_0,y_0)$ determined at the chosen fixed $z$.

Using eqs.~\ref{eq:cal1} and \ref{eq:cal2},  eqs.~\ref{eq:alpha} and \ref{eq:phi} become
\begin{equation} \label{eq:alpha_no_off}
\alpha (x,y)=\alpha_0 \left(  \frac{\Delta V_0} {\Delta V(x,y)}\right)
\end{equation}
and
\begin{equation} \label{eq:phi_no_off}
\Phi^*(x,y)=-\alpha_0 \left(  \frac{ V^-(x,y)}{\Delta V(x,y)}\Delta V_0- V^-_0\right)
\end{equation}

\noindent where $\Delta V_0=V^+_0-V^-_0$. According to eqs.~\ref{eq:alpha_no_off} and \ref{eq:phi_no_off}, both the gating efficiency $\alpha$ and the potential $\Phi^*$ can be fully expressed in terms of measurable quantities, up to a common scaling factor $\alpha_0$.

\section{Measurement of $\Phi^*(x,y,z)$}

Eq.~\ref{eq:phi_no_off} can be directly applied to the measurement of $\Phi^*$ at an arbitrary location $(x,y,z)$ if the reference data $V^-_{\rm{0}}(z)$ and $V^+_{\rm{0}}(z)$ are available. Since the scaling factor $\alpha_0$ in eq.~\ref{eq:phi_no_off} is generally $z$-dependent, we obtain

\begin{equation} \label{eq:phi_no_offxx}
\Phi^*(x,y,z)=-\alpha_0(z) \left(\frac{V^-(x,y,z)}{\Delta V(x,y,z)}\Delta V_{0}(z)-V^-_0(z)\right).
\end{equation}

\noindent If we assume the junction geometry to be that of a plate capacitor, whence the potential drops linearly between the electrodes and therefore $\alpha_0(z)=d/(d+z)$ (for definitions of $d$, $z$ and $z_{\rm{tip}}$ see Fig.~2a of the supplement), we finally obtain

\begin{equation} \label{eq:phi_no_offxxx}
\Phi^*(x,y,z)=-\frac{d}{d+z} \left(\frac{V^-(x,y,z)}{\Delta V(x,y,z)}\Delta V_{0}(z)-V^-_0(z)\right)
\end{equation}


\begin{thebibliography}{99}
\bibitem{DTFnote}
The DFT calculation was carried using the SIESTA package \cite{Siesta1,Siesta2}. The exchange-correlation part was treated within the local-density approximation which gives realistic results for strongly bound adsobates like the Ag adatom considered here. The adatom adsorbs at the hollow site with a height of $2.11$~\AA, which is $0.25$~\AA~closer to the surface than its ideal position would be. In order to calculate the electrostatic potential in real space,
we decompose the charge density into two parts, by the clean surface and by the adatom. The main contribution of the Ag adatom originates from a region of about $4$~\AA~around the adatom. Since the potential of the clean neutral surface decays exponentially with increasing height, only the charge density contribution due to the adatom causes a measurable electrostatic potential at the position of the QD. This potential is evaluated from the charge density by solving Poisson's equation in real space, without any supercell periodicity.
\bibitem{screening}
The screening was evaluated by considering an elementary electrostatic model of a point charge in front of an infinite metal surface.
\end{thebibliography}
\end{document}